\documentclass[]{interact}

\usepackage{epstopdf}
\usepackage[caption=false]{subfig}

\usepackage[longnamesfirst,sort]{natbib}
\bibpunct[, ]{(}{)}{;}{a}{,}{,}

\usepackage{graphicx}
\usepackage{moreverb,url}

\usepackage[dvipsnames]{xcolor}
\usepackage{enumitem}

\usepackage{amsmath}
\usepackage{amssymb}
\usepackage{tabularx}

\usepackage{float}

\usepackage[hidelinks]{hyperref}

\usepackage{etoolbox}
\AtBeginEnvironment{thebibliography}{%
  \interlinepenalty=10000
}

\makeatletter
\renewcommand\@maketitle{%
  \bgroup
    \parindent0pt
    \vspace*{36pt}
    {\articletypefont{\@articletype}\par}%
    \vskip13pt
    {\titlefont{\@title}\par}%
    \vskip13pt
    {\authorfont\@author\par}%
    \ifsuppldata\else
    \vskip17pt\vspace{1cm}%  <-- added 1cm here
    {\receivedfont{\bfseries ARTICLE HISTORY\\}\@received\par}%
    \fi
    \vskip13pt
  \egroup}
\makeatother

\begin{document}

\title{Methods for Pitch Analysis in Contemporary Popular Music: Multiphonic Tones Across Genres}

\author{
\name{Emmanuel Deruty\textsuperscript{a,b}\thanks{CONTACT Emmanuel Deruty. Email: emmanuel.deruty@sony.com}, David Meredith\textsuperscript{b}, Yann Mac\'e\textsuperscript{c}, Luc Leroy\textsuperscript{c}, Dima Tsypkin\textsuperscript{d} and Pascal Arbez-Nicolas\textsuperscript{e}}
\affil{\textsuperscript{a}Sony Computer Science Laboratories, 6 rue Amyot, 75005 Paris, France; \textsuperscript{b}Department of Architecture, Design and Media Technology, Aalborg University, Rendsburggade 14, 9000 Aalborg, Denmark; \textsuperscript{c}Neodrome Entertainment, 5 rue Vernet, 75008 Paris, France; \textsuperscript{d}Independent performing musician, Senjak, Belgrade, Serbia; \textsuperscript{e}Citizen Records, 2 rue Jacques Cellerier, 21000 Dijon, France}
}

\maketitle

%%%%%%%%%%%%%%%%%%%%%%%%%%%%%%%%%%%%%%%%%%%%%%%%%%%%%%%%%%%%%%%%%%%%%%%%%%%%%%%%
% Abstract
%%%%%%%%%%%%%%%%%%%%%%%%%%%%%%%%%%%%%%%%%%%%%%%%%%%%%%%%%%%%%%%%%%%%%%%%%%%%%%%%

\vspace{1cm}
\begin{abstract}

%This study compares tones referred to as ``multiphonics'' in the contemporary Western classical tradition with electronic tones drawn from commercial contemporary popular music, some belonging to frequently used types. Using listening tests and signal analysis, it examines how, in both cases, tone-design practices give rise to multiple and ambiguous pitch percepts. By relating listeners' perceived pitches to spectral and temporal features of the signal, the study shows how popular music producers use techniques similar to those of contemporary classical composers, resulting in variable pitch perceptions and thereby extending traditional notions of pitch in modern music production.

This study argues that electronic tones routinely used in contemporary popular music — including 808-style bass and power chords — are structurally and perceptually equivalent to \textit{multiphonics} in contemporary classical music. Using listening tests (n=10) and signal analysis, we show that both types of tones elicit multiple, listener-dependent pitch percepts arising from similar spectral and temporal features. These findings suggest that pitch ambiguity is not confined to experimental classical contexts but is also a feature of mainstream music production.

\end{abstract}


%%%%%%%%%%%%%%%%%%%%%%%%%%%%%%%%%%%%%%%%%%%%%%%%%%%%%%%%%%%%%%%%%%%%%%%%%%%%%%%%
% Main Content Start
%%%%%%%%%%%%%%%%%%%%%%%%%%%%%%%%%%%%%%%%%%%%%%%%%%%%%%%%%%%%%%%%%%%%%%%%%%%%%%%%

\newpage
\section{Introduction}\label{sec:introduction}

In the Western classical tradition, \textit{multiphonics} are tones produced by a normally monophonic source in which two or more pitches are heard simultaneously \citep{campbell2001multiphonics}. The term \textit{pure multiphonics} refers to cases in which the perceived pitches align with harmonic partials of the fundamental \citep{walter2020multiphonics}. Such techniques have been documented on the oboe \citep{bartolozzi1982woodwinds}, cello \citep{fallowfield2019cello}, horn \citep{mikulka2018practical}, and piano \citep{walter2020multiphonics}.

Tones used in contemporary popular music -- including widespread TR-808– style basses and single harmonic tones resulting from power chords -- have been shown to elicit multiple pitch percepts \citep{deruty2025vitalictemperament,deruty2025primaal}, which vary across listeners \citep{deruty2025multiple}. Using listening tests and signal analysis, the present study shows that these tones exhibit structural and perceptual properties equivalent to \textit{multiphonics} in contemporary Western classical music. This result is significant, as it extends the relevance of multiphonics to music with a much broader audience \citep{statista2018genres}.

While \citet{deruty2025multiple} demonstrates that different listeners can perceive different sets of pitches from sequences of quasi-harmonic electronic tones in popular music, the present study makes three additional contributions: (a) the inclusion of Western contemporary classical material alongside electronic tones; (b) a focus on isolated tones rather than sequences; and (c) a comparison of human pitch perception with the output of monophonic pitch trackers.

The study incorporates music and direct input from the electronic musician Vitalic\footnote{https://www.vitalic.org/} and the Hyper Music production company\footnote{https://www.hyper-music.com/}, both of whom validated the findings from a music producer's perspective.

The study's findings bear on assumptions in Music Information Retrieval, where harmonic complex tones are treated as conveying a single pitch \citep{kim2018crepe,riou2025pesto}. The pitch-tracker comparison illustrates how algorithmic output both diverges from and converges with human perception.

Section~\ref{sec:background} defines the terminology used in the study, while Section~\ref{sec:methods} describes the employed methods. Sections~\ref{sec:single} to~\ref{sec:inharm} examine the relationship between four types of increasingly atypical tones and their perceived pitches: quasi-harmonic tones designed to convey a single pitch (Section~\ref{sec:single}); quasi-harmonic tones designed to convey several specified pitches (Section~\ref{sec:multiphonics_specified}); quasi-harmonic tones designed to convey several unspecified pitches (Section~\ref{sec:multiphonics_unspecified}); and inharmonic tones (Section~\ref{sec:inharm}). Supplementary material containing the audio samples, listener profiles, listening conditions, and raw listening test results is available at:

\url{https://vnht-suppl-mat.s3.eu-west-3.amazonaws.com/index.html}

\newpage
\section{Terminology}\label{sec:background}

\subsection{Partials, harmonics, and harmonic complex tones}

The study borrows terminology from \citet{deruty2025multiple}. A \textit{partial} is defined as the spectral representation of a single periodic sine wave. A \textit{harmonic complex tone} is an ensemble of partials whose frequencies are integer multiples of a fundamental frequency ($f_0$). An \emph{overtone} or \emph{upper partial} is any partial in a tone other than $f_0$. A \textit{harmonic} is a sinusoidal component (partial) of a harmonic tone, where the $n^{\text{th}}$ harmonic has a frequency equal to $n$ times the fundamental. \textit{Inharmonicity} refers to a discrete inharmonic partial configuration in which overtone ratios deviate from integer multiples of a base frequency.

\subsection{Quasi-harmonicity}

No tone in this study possesses partials that strictly follow a harmonic relationship. However, degrees of inharmonicity may vary, allowing the identification of \emph{quasi-harmonic tones}, in which each partial of the least-deviating harmonic series can still be associated with a partial of the original tone. In such cases, $f_0$ may be defined as the fundamental of this least-deviating series \citep{rasch1982perception}, and the $n^{\text{th}}$ harmonic's frequency lies near the $n$-fold multiple of the $f_0$ of the least-deviating harmonic series. In this study, a tone is classified as \textit{inharmonic} when the association between its upper partials and the theoretical harmonic positions becomes difficult (e.g., Section~\ref{sec:piano}), or when the upper partials show regular spacing but the inferred $f_0$ does not fit (e.g., Sections~\ref{sec:silver} and~\ref{sec:nofun}).

\subsection{Spectral and temporal modeling}\label{subsec:modeling}

Yost \citeyearpar{yost2009pitch} distinguishes between \textit{spectral} and \textit{temporal} models of pitch perception. In spectral modeling, pitch is extracted from spectral components that are \textit{resolved} by the auditory periphery -- that is, components whose frequency spacing exceeds the width of critical bands. While temporal modeling typically derives pitch from the highest autocorrelation peak beyond zero lag \citep{rabiner1976comparative}, we also consider frequency differences between adjacent partials as indicators of temporal structure \citep{deruty2025vitalictemperament,deruty2025multiple}. Depending on the tone, the spectral and temporal approaches may either converge or diverge.

\section{Methods}\label{sec:methods}

\subsection{List of samples}\label{sec:listofsamples}

Table~\ref{table:list} lists the samples used in the study (all \textit{single tones}) sorted by increasing average number of perceived pitches. This measure is derived from the number of pitches reported during the listening tests, weighted by their certainty ratings -- see Section~\ref{sec:listeningtests} and the supplementary material. 

\begin{table*}[h!]
\centering
\small
\setlength{\tabcolsep}{4pt}
\begin{tabularx}{\textwidth}{l l l l X}
\hline
\textbf{Sample} & \textbf{Section} & \textbf{Production} & \textbf{Source} & \textbf{Number}
\\ & & \textbf{method} &  & \textbf{of perceived}
\\ & & & & \textbf{pitches}  \\
\hline
Cello F2 tone          & \ref{sec:single} & Acoustic & This study & 1.64 \\
Cello natural harmonic         & \ref{sec:cellomultiphonics} & Acoustic & This study & 1.66 \\
Horn multiphonic       & \ref{sec:Weber} & Acoustic & Weber \citeyearpar{weber2012london} & 1.86 \\
Oboe multiphonic       & \ref{sec:holliger} & Acoustic   & Holliger \citeyearpar{holliger1971studie} & 1.95 \\
Cello multiphonic (1)  & \ref{sec:cellomultiphonics} & Acoustic & This study & 2.14 \\
808-type bass          & \ref{sec:silver} & Electronic & Primaal \citeyearpar{primaal2023silver} & 2.14 \\
Inharmonic synth.      & \ref{sec:nofun} & Electronic & Vitalic \citeyearpar{vitalic2005nofun} & 2.20\\
Power chord            & \ref{sec:powerchord} & Electronic & Metallica \citeyearpar{metallica1988justice} & 2.39 \\
FM bass + distortion            & \ref{sec:seethesea} & Electronic & Vitalic \citeyearpar{vitalic2009seetheseared} & 2.48 \\
FM bass + filter bank         & \ref{sec:fire} & Electronic & Primaal \citeyearpar{primaal2023fire} & 2.52 \\
Cello multiphonic (2)  & \ref{sec:cellomultiphonics} & Acoustic & Fallowfield \citeyearpar{fallowfield2019cello} & 2.56 \\
Piano multiphonic      & \ref{sec:piano} & Acoustic & Walter \citeyearpar{walter2020multiphonics} & 2.64 \\

\hline
\end{tabularx}
\vspace{0.3cm}
\caption{Samples used in the study.}
\label{table:list}
\end{table*}

\newpage

The numbers of perceived pitches are commensurable with those reported in \citet{deruty2025multiple}: no tone transmits a single pitch -- even the ``normal'' cello tone -- and the most complex tones transmit up to 2.6 pitches. Although these results should be interpreted as outcomes of \textit{analytical listening} on \textit{isolated samples}, they nevertheless contribute to the debate concerning the auditory system behaving as a frequency analyzer \citep{dixonward1970, plomp1976aspects, turner1977ohm}.

A significant observation is that electronically produced tones from commercially successful popular music yield numbers of perceived pitches comparable to those found in contemporary classical music multiphonics.

\subsection{Listening tests}\label{sec:listeningtests}

\subsubsection{Procedure.} Ten musically trained listeners participated in the listening tests, using a piano to match the pitches they perceived in twelve samples. Although this setup departs from standard pitch-matching procedures -- e.g.,  \citep{shackleton1994role}, it was adopted because it is familiar to trained musicians as a \textit{musical-dictation task} -- transcribing the pitches they hear. Using isolated samples rather than sequences elicits a more analytical listening mode than in \citet{deruty2025multiple}. For each sample, listeners were asked to report all perceived pitches, together with a degree of certainty between zero and one, and to indicate whether each pitch sounded out of tune (too low or too high).

%\vspace{-.25cm}

\subsubsection{Uncontrolled listening conditions.}

The listening environment was intentionally left uncontrolled. Indeed, one aim of controlled settings is reproducibility, and the present study instead aims to make the following conclusions reproducible \textit{regardless of listening conditions}: (a) a variety of tones (from quasi-harmonic to inharmonic), when presented as a musical-dictation task, elicit different perceived pitches across listeners; (b) the perceived pitches are generally harmonically related; and (c) tones typically theorized to convey a single pitch may behave similarly.

Sanity checks were conducted to ensure that listening conditions were not the primary determinant of pitch perception. For example, although the mean number of perceived pitches (certainty-weighted) across all listeners and samples was 2.2, listener~8 reported an average of 3.8 pitches, whereas listener~6 reported only 1.3, despite both using the same monitoring system in the same room. Also, listener 7 frequently perceived lower frequencies than all others while using the same system as listeners 2, 3, 6, and 8.

\subsubsection{Pitch trackers.} Listening test results are compared with analysis outputs from CREPE \citep{kim2018crepe} and PESTO \citep{riou2025pesto}. Although both are monophonic, their outputs show multiple predicted pitch values -- \textit{pitch jumps} \citep{watkins2024revisiting}, with the windowed estimates being aggregated across the entire duration of the sample.

%\vspace{-.1cm}

\subsection{Signal- and listening-based comparative interpretation}\label{sec:signal}

Sections~\ref{sec:single} to~\ref{sec:inharm} report the pitches perceived during the listening tests. When possible, these pitches are associated with (a) the tone’s fundamental frequency -- obtained through spectral and/or temporal modeling -- and (b) its upper partials. In many cases (c), pitches cannot be linked to specific partials but only to octave transpositions of partials, reflecting listener reports -- also noted in \citet{deruty2025multiple} -- of the frequent difficulty in determining the octave associated with a pitch class. In some instances, multiple plausible associations exist, in which case the choice is necessarily somewhat arbitrary. Such interpretation becomes increasingly challenging as tones grow more inharmonic.

To clarify the representations, the study uses an \textit{ad hoc} distance $d$ from the fundamental, whereby associations of type (a) have distance~0 and those of type (b) have distance~1. Octave transpositions (c) increase the distance by one. For example, a pitch that does not correspond to a partial but whose \textit{pitch class} matches a salient upper partial will result in $d=2$.

\subsection{Psychoacoustic weighting}\label{subsec:iso}

Signal analyses in this study incorporate equal-loudness–contour weighting, recognising that humans are not equally sensitive to all frequencies \citep{fletcher1933loudness}. Several models exist for weighting the power spectrum to better reflect perception \citep{fletcher1933loudness,robinson1956re,skovenborg2004evaluation}. One such model is ISO226:2023 \citep{iso2262023}. Following \citet{deruty2025vitalictemperament,deruty2025multiple}, we apply the 50-phon equal-loudness contour from ISO226:2023 prior to signal analysis so that the results more closely reflect what listeners actually hear. In the figures, this is indicated by the label `weighted signal'.

\newpage

\subsection{Involvement of music producers}

The study involves the participation of the electronic musician Vitalic and the Hyper Music production company, both of whom provided insights on the use of multiphonic tones in contemporary popular music. Vitalic is the stage name of Pascal Arbez-Nicolas, whose career is documented by Phares \citeyearpar{phares2024vitalic}. A distinctive feature of his production style is what he describes as ``a melody inside the bass'' \citep{musicradartech2014vitalicITV}, referring to a second melodic line embedded within sequences of multiphonics \citep{deruty2025vitalictemperament, deruty2025multiple}. The Hyper Music production company \citep{deruty2022melatonin, deruty2022development} is directed by Yann Mac\'e and Luc Leroy. Under its \textit{Primaal} brand, it produces music for mainstream advertising that routinely employs pure multiphonics -- particularly, though not exclusively, in bass lines \citep{deruty2025multiple, deruty2025primaal}.

%% %%%%%%%%%%%%%%%%%%%%%%%%%%%%%%%%%%%%%
%% Section
%% %%%%%%%%%%%%%%%%%%%%%%%%%%%%%%%%%%%%%

%\vspace{cm}

\section{A quasi-harmonic tone designed to convey one pitch}\label{sec:single}

Western classical instruments produce quasi-harmonic complex tones. In such tones, the harmonics' frequencies are near-multiples of the fundamental, and the pitch predicted by temporal modeling corresponds to the $f_0$. A debate has persisted for centuries over how many harmonics in a complex tone can be ``heard out'' as having their own pitches \citep{helmoltz1885sensations,plomp1964ear,turner1977ohm}. The current consensus is that Western musical tones transmit a single pitch \citep{dixonward1970,plomp1976aspects}.

\begin{figure}[h!]
\includegraphics[width=\textwidth]{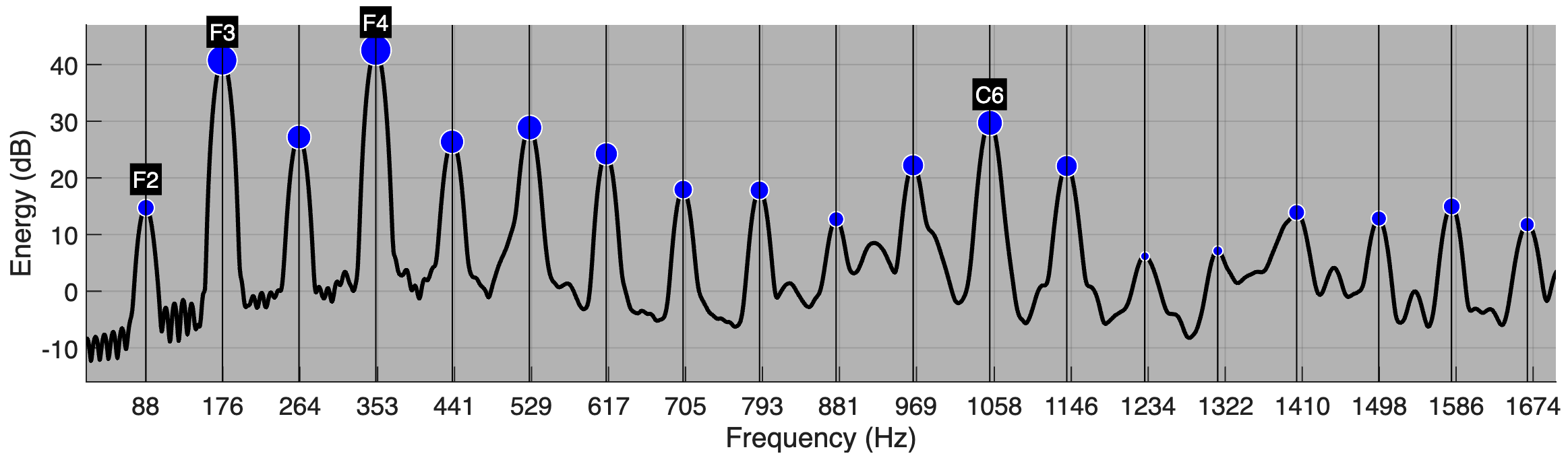}
\caption{Cello tone, F2 on the C string, weighted audio. Power spectrum. The dot's size reflects the energy. The x-axis shows multiples of the $f_0$.} \label{fig:cello}
\end{figure}

\begin{figure}[h!]
\includegraphics[width=\textwidth]{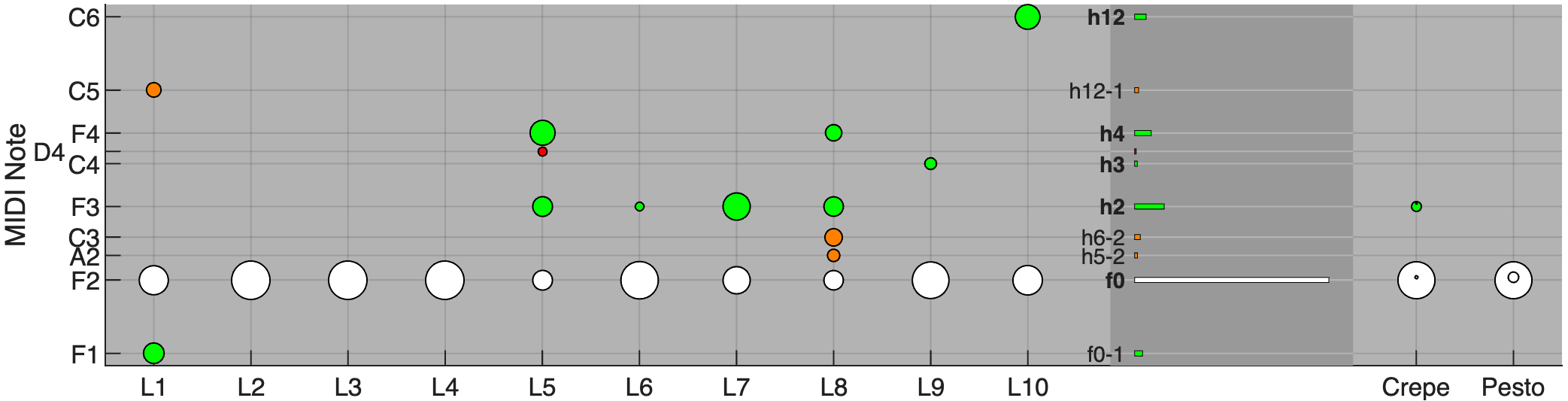}
\caption{Cello tone, F2 on the C string, listening test results. In white, pitches for which $d=0$; in green, $d=1$; in orange, $d=2$; in red, no association with the signal. The same protocol will be used in the subsequent figures. The bar graph to the left of the pitch-tracker results compiles the perceived pitches together with their associations. The label ``h5-2'', for example, indicates an association with harmonic~5, two octaves below. Here, C5 is not associated with harm.~6 but is described as one octave below C6 (harm.~12), as the C6 partial is salient in Fig.~\ref{fig:cello}.} \label{fig:cello_results}
\end{figure}

In this study, we use a F2 monophonic cello sample as a \textit{control}. According to Fig.~\ref{fig:cello}, the tone is slightly inharmonic, and the $f_0$ is not the loudest partial. As in the following sections, any indicated pitch label (e.g., ``B1'') should be understood as an approximation in equal temperament, except when otherwise specified. Fig.~\ref{fig:cello_results} shows the listening test results for this sample. The $f_0$ F2 yields a strong consensus, with F3 (harm.~2) and F4 (harm.~5) also perceived by more than one listener. CREPE's distribution reflects a few octave jumps between F2 and F3.

\vspace{.5cm}

%% %%%%%%%%%%%%%%%%%%%%%%%%%%%%
%% Specified pitches
%% %%%%%%%%%%%%%%%%%%%%%%%%%%%%

\newpage

\section{Quasi-harmonic tones designed to convey several specified pitches}\label{sec:multiphonics_specified}

This section concerns tones that are quasi-harmonic according to the criteria in Section~\ref{sec:background}, and for which the pitches that the listener is intended to hear are specified, e.g., by a musical score.

\subsection{Cello multiphonics}\label{sec:cellomultiphonics}

Pure multiphonics on the cello have been studied by Helen Fallowfield \citeyearpar{fallowfield2019cello}. Pitches above the fundamental are obtained by applying pressure at nodal points along the string. Fallowfield focuses on multiphonics intended to transmit \textit{three or more pitches} by inducing harmonics from neighbouring nodes to sound simultaneously. To complement such examples, we include consideration of cello tones (played and recorded by two of the authors) intended to transmit \textit{only two pitches}. 

\vspace{.2cm}

\begin{figure}[h!]
\includegraphics[width=\textwidth]{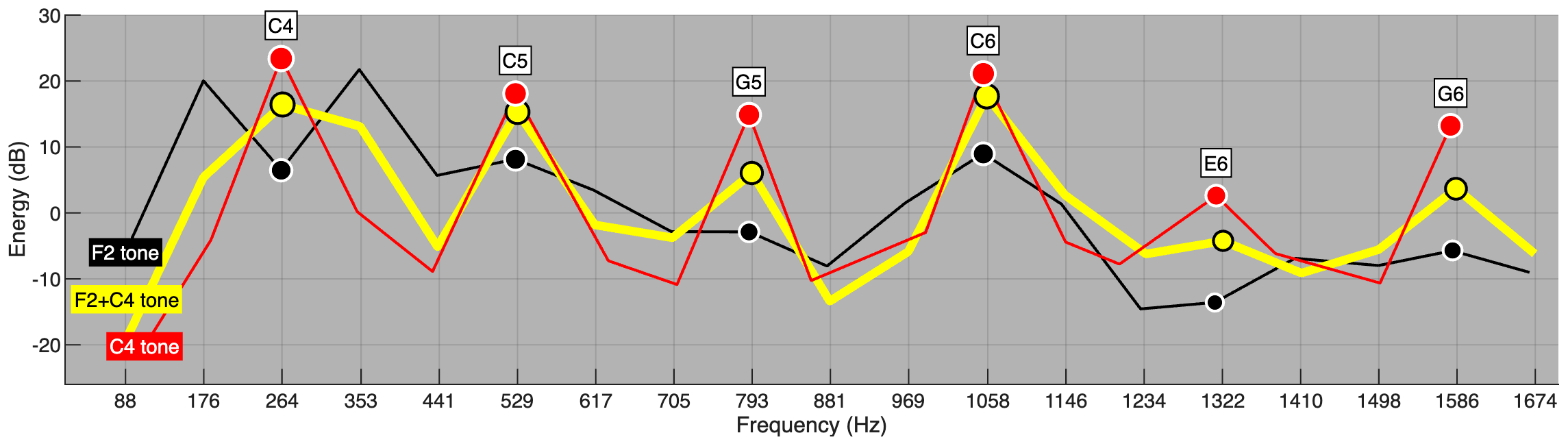}
\caption{Cello recordings, weighted audio. Power spectrum peaks corresponding to F2 (fundamental), F2+C4 (multiphonic), and C4 (harmonic 3). The dots highlight the partials corresponding to C4. The x-axis shows multiples of the F2 $f_0$.} \label{fig:dima}
\end{figure}

\begin{figure}[h!]
\includegraphics[width=\textwidth]{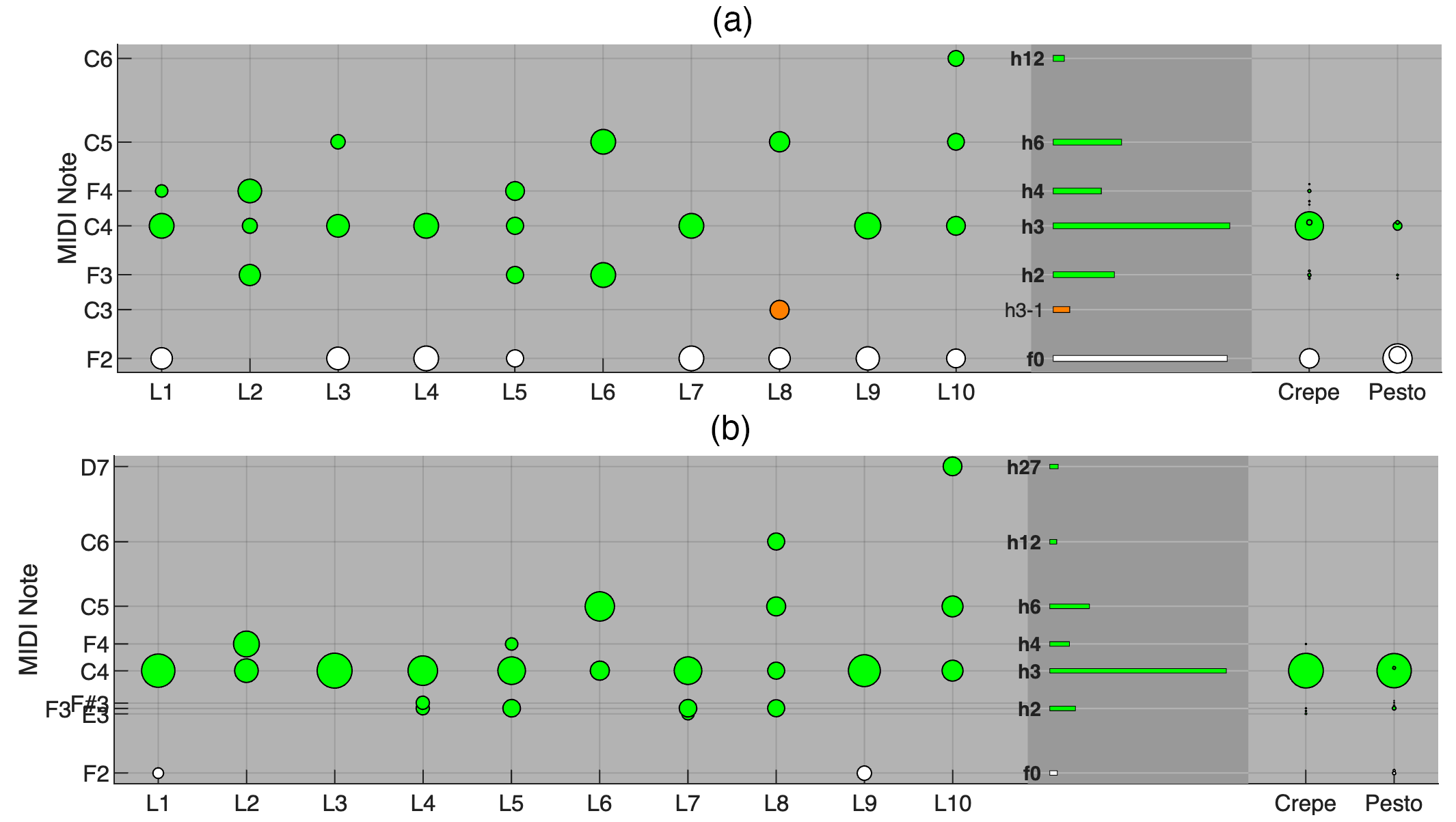}
\caption{(a) F2+C4 and (b) C4 tones from Fig.~\ref{fig:dima}, listening test results. In (b), ``h27'', the 27\textsuperscript{th} harmonic of F2, is the 9\textsuperscript{th} harmonic of C4.} 
\label{fig:dima_results}
\end{figure}

%\vspace{-1cm}

\newpage

\subsubsection{Only two pitches.} Fig.~\ref{fig:dima} compares the power-spectrum peaks of the F2 tone from Fig.~\ref{fig:cello}, a multiphonic F2--C4 tone ($f_0$ + harm. 3), and harmonic~3 alone. The analysis suggests that the multiphonic nature of the F2--C4 arises from a balance between the energy of the partials corresponding to the upper tone(s) and the energy of all partials.

Fig.~\ref{fig:dima_results} shows the listening test results for these two samples. The F2+C4 tone in (a) yields a consistent perception of multiple harmonics, with emphasis on the two intended pitches. The consistent perception of C4, corresponding to a subset of louder partials in harmonic relations, is reminiscent of Fallowfield's observation that the ear tends to organise multiphonic partials into groups corresponding to distinct harmonic series \citep{fallowfield2019cello}. CREPE's distribution is the result of pitch jumps, recalling observations according to which such jumps may reflect signal features instead of deriving from technical defects \citep{watkins2024revisiting}. The C4 harmonic in (b) leads to a more consensual perception, although several other harmonics are also consistently perceived as pitch.

\subsubsection{Three pitches and more.}

\begin{figure}[h!]
\includegraphics[width=\textwidth]{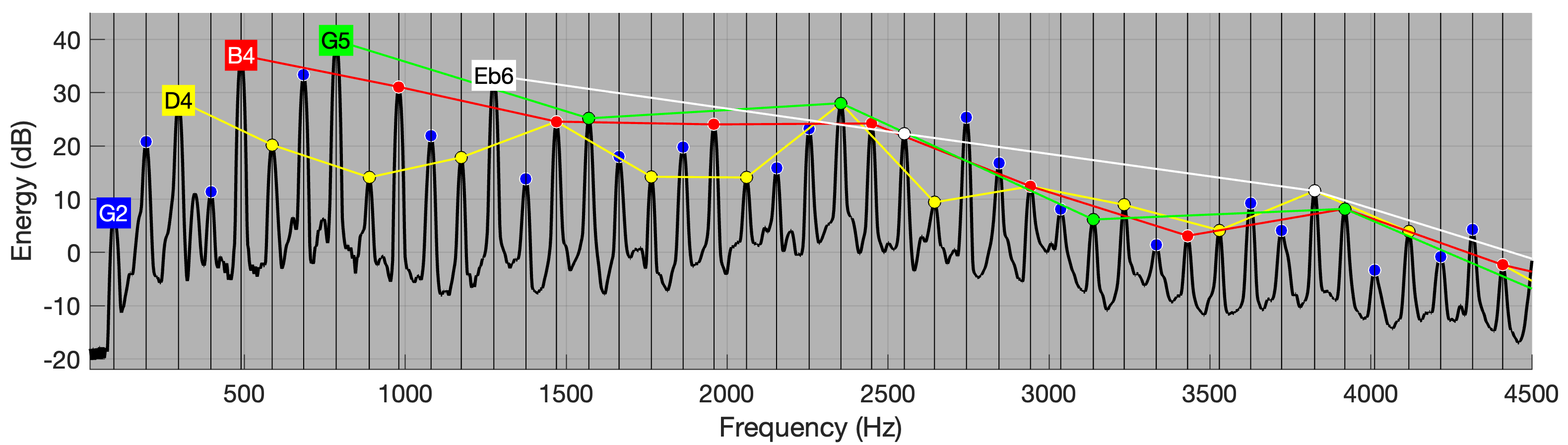}
\caption{Fallowfield's ``AudioEx11 III 3–5–8–13 EK'' sample \citep{fallowfield2019cello}, weighted audio. Power spectrum. Harmonic tones with fundamentals corresponding to partials 1, 3, 5, 8, and 13 are shown in blue, yellow, red, green, and white, respectively.} \label{fig:fallowfield}
\end{figure}

Fig.~\ref{fig:fallowfield} illustrates the analysis of a pure multiphonic tone provided by Fallowfield \citeyearpar{fallowfield2019cello} with G2 as the fundamental, in which the intended transmitted pitches correspond to the 3\textsuperscript{rd}, 5\textsuperscript{th}, 8\textsuperscript{th}, and 13\textsuperscript{th} partials. The first partial (fundamental) is relatively weak. The pitches intended to be transmitted appear to correspond to groups of salient partials. 

\begin{figure}[h!]
\includegraphics[width=\textwidth]{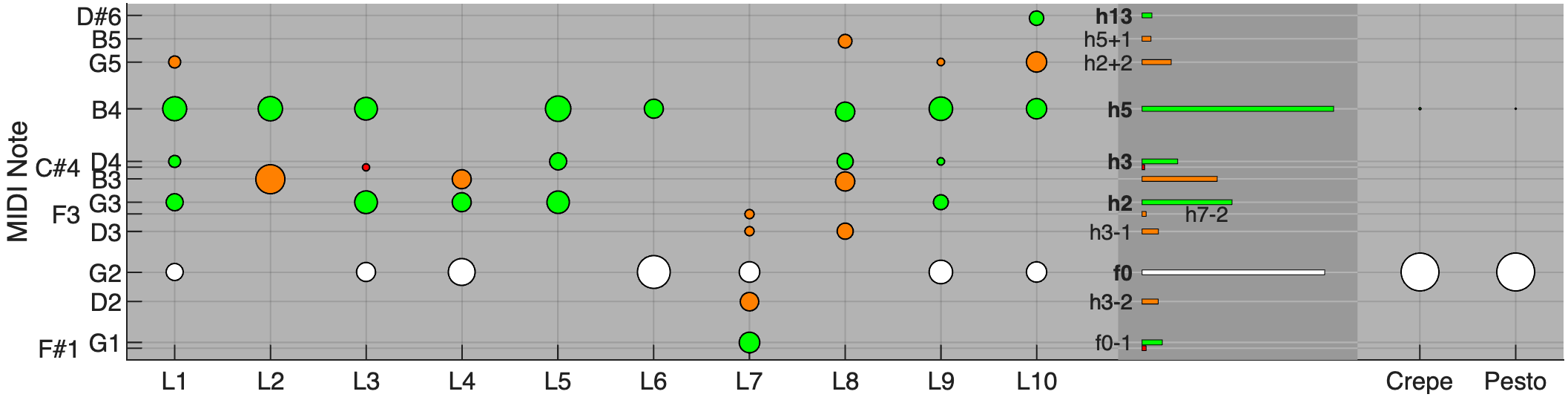}
\caption{Fallowfield's tone from Fig.~\ref{fig:fallowfield}, listening test results.} \label{fig:fallowfield_results}
\end{figure}

Fig.~\ref{fig:fallowfield_results} shows the listening test results for this sample. The transmitted pitches do not correspond to the intended ones. They mostly align with harm. 5, the $f_0$ and some other harmonics, as well as octave transpositions of harmonics. Both PESTO and CREPE reflect the pitch predicted by temporal modeling.

\newpage
\subsection{Weber, \textit{Concertino} for horn and orchestra}\label{sec:Weber}

Weber's \textit{Concertino} in E, Op.~45, was written in 1806 and 1815 \citep{warrack1976carl}. Bars 168--174 include horn multiphonics \citep{weber1806concertino}. According to Kirby \citeyearpar{kirby1925horn}, the chords actually played differ from those notated, suggesting that the written notation is approximate. Kirby proposes that playing such chords may be done using combination tones \citep{helmoltz1885sensations} generated by the interaction of a played tone with an additional sung tone whose perceived pitch is close to one of the harmonics of the played tone.

\vspace{.3cm}

\begin{figure}[h!]
\includegraphics[width=\textwidth]{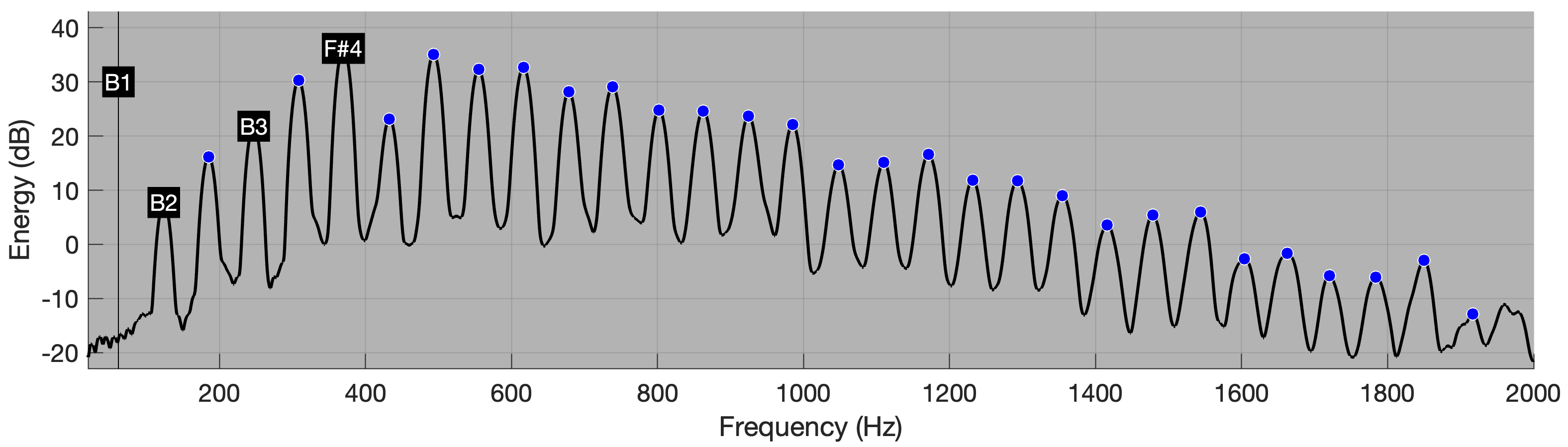}
\caption{Weber, \textit{Concertino} in E, bar 172 multiphonic, weighted audio. Power spectrum. The vertical line shows the predicted pitch value according to temporal modeling (B1).} \label{fig:weber}
\end{figure}

\newpage

Fig.~\ref{fig:weber} shows the tone's architecture for bar 172, derived from a 2012 Chandos recording \citep{weber2012london}. The notes indicated on the score are B1-B3-F\#4. Kirby notes that the multiple perceived pitches do not correspond to equal temperament, which is consistent with \citet{deruty2025multiple}. For this sample, the perceived pitches only partially match the specified ones. They include those corresponding to the $f_0$ as predicted by temporal modeling, to upper harmonics -- especially B2 (harm. 2) -- and to a few octave transpositions. Although the F\#4 partial is the loudest, it does not yield a consistently perceived pitch, suggesting that pitch perception arises from subsets of partials rather than from individual components. Despite some octave jumps, CREPE and PESTO generally follow the (missing) $f_0$.

\begin{figure}[h!]
\includegraphics[width=\textwidth]{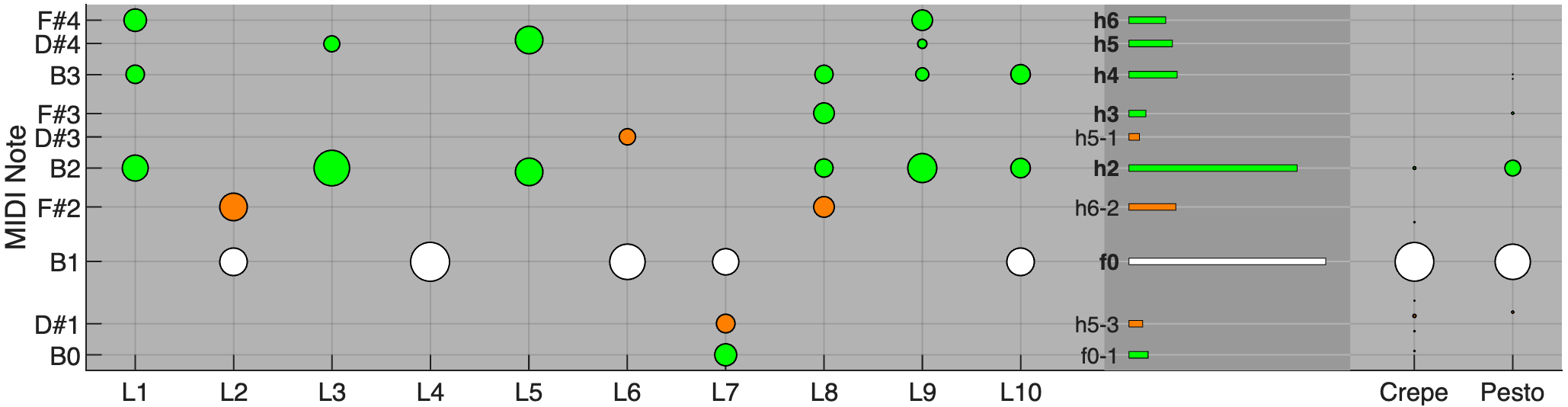}
\caption{Weber, \textit{Concertino}, tone from Fig.~\ref{fig:weber}, listening test results. The association between D\#1 and harmonic 5 is doubtful.} \label{fig:weber_results}
\end{figure}

%% %%%%%%%%%%%%%%%%%%%%%%%%%%%%
%% Unspecified pitches
%% %%%%%%%%%%%%%%%%%%%%%%%%%%%%

\section{Quasi-harmonic tones designed to convey several unspecified pitches}\label{sec:multiphonics_unspecified}

In this section, we consider tones that are quasi-harmonic according to the criteria in Section~\ref{sec:background}, and for which no pitches to be transmitted are specified; only the instrumental means of producing the tones are available.

\subsection{Guitar power chord, in \textit{...And Justice for All}}\label{sec:powerchord}

\begin{figure}[h!]
\includegraphics[width=\textwidth]{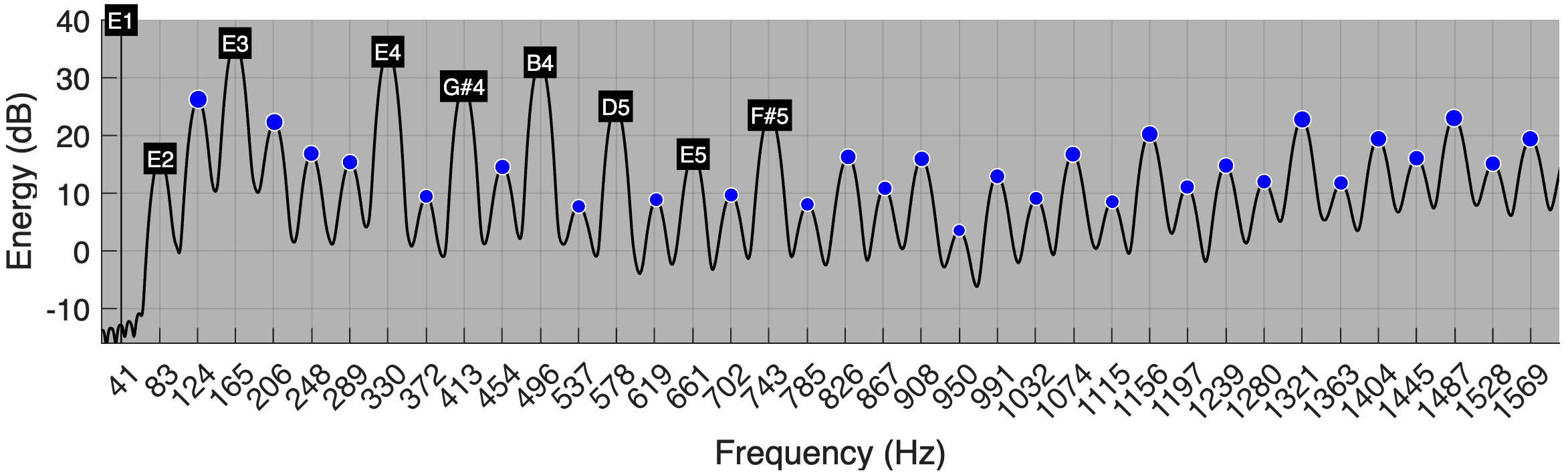}
\caption{Metallica, \textit{... And Justice for All}, 1'07--1'14, extracted guitar track (one power chord), weighted audio. Power spectrum. The vertical line shows the predicted pitch value according to temporal modeling (E1). The x-axis shows multiples of the $f_0$.} \label{fig:pchord}
\end{figure}

Playing a power chord results in a single multiphonic harmonic complex tone \citep{deruty2025multiple}. The notes played on the instrument are separated by a fifth and, in some cases, an octave. Intermodulation distortion generates combination tones from the original partials \citep{newell2017recording}. These three characteristics make power chords similar to Weber's horn multiphonics. A previous listening test \citep{deruty2025multiple} examined the perception of pitches in a power chord produced from isolated guitar samples. Here, we consider a power chord from Metallica's \textit{... And Justice for All} \citep{metallica1988justice}, extracted through source separation using the X-UMX algorithm \citep{sawata2021all}. Fig.~\ref{fig:pchord} shows the result as being a harmonic tone with a missing fundamental.

\begin{figure}[h!]
\includegraphics[width=\textwidth]{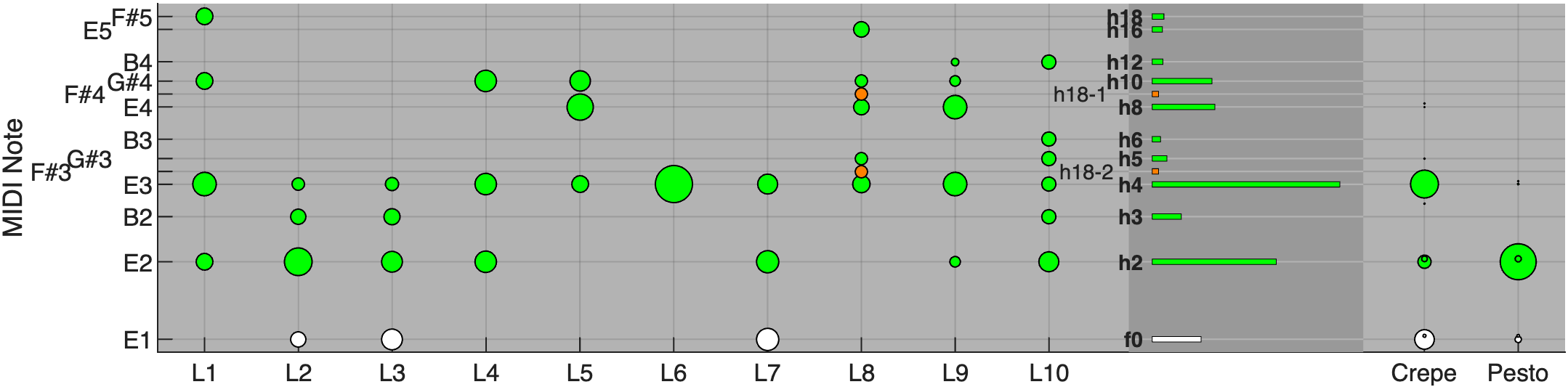}
\caption{Metallica, \textit{... And Justice for All}, tone from Fig.~\ref{fig:pchord}, listening test results.} \label{fig:pchord_results}
\end{figure}

%\newpage

Fig.~\ref{fig:pchord_results} shows the listening test results for this sample. The span of perceived pitches is wide and generally corresponds to harmonics -- including high-order harmonics, which suggests that in this case, some pitches corresponding to harmonics above the 8\textsuperscript{th} can be resolved, contradicting \citet{plomp1976aspects}. CREPE features octave jumps, while PESTO focuses on harm. 2.

\subsection{Vitalic, \textit{See the Sea (Red)}, bass}\label{sec:seethesea}

\begin{figure}[h!]
\includegraphics[width=\textwidth]{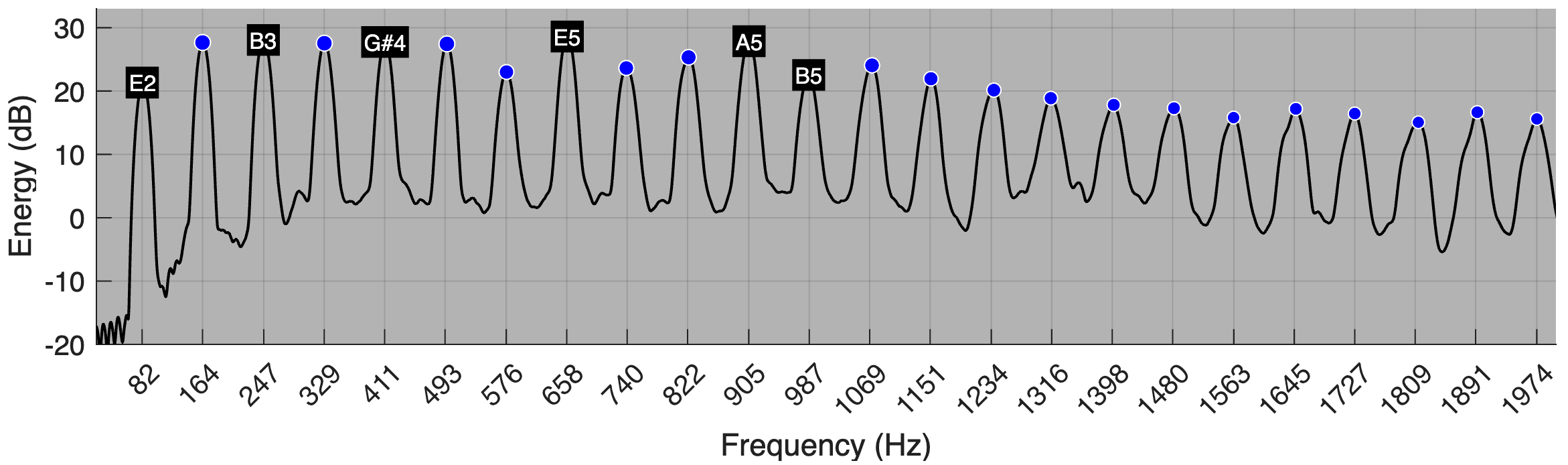}
\caption{Vitalic, \textit{See the Sea (red)}, 0'50--0'52, extracted bass track, weighted audio. Power spectrum. The x-axis shows multiples of the $f_0$.} \label{fig:seethesea}
\end{figure}

This section focuses on a tone from Vitalic's \textit{See the Sea (red)} \citep{vitalic2009seetheseared}, produced using Native Instruments' FM8 with additional non-linear distortion -- obtained through source separation. According to Vitalic, such tones are specifically \textit{designed} to elicit ambiguous pitch perception, varying in particular with the listening system and the presence of other tracks. Fig.~\ref{fig:seethesea} reveals a harmonic tone with several equally loud partials.

\vspace{.2cm}

\newpage

\begin{figure}[h!]
\includegraphics[width=\textwidth]{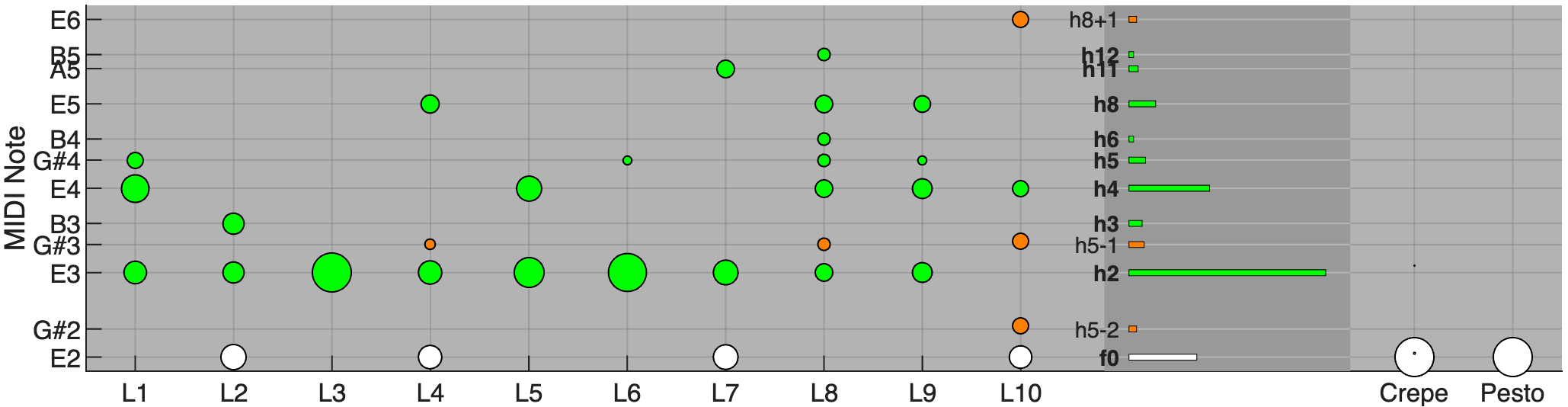}
\caption{Vitalic, \textit{See the Sea (red)}, tone from Fig.~\ref{fig:seethesea}, listening test results.} \label{fig:seethesea_results}
\end{figure}

Fig.~\ref{fig:seethesea_results} shows the listening test results for this sample. Despite a loud fundamental (E2) driving the outputs of CREPE and PESTO, the most frequently perceived pitch is harm.~2 (E3). Other perceived pitches correspond to a range of harmonics, including harmonics~11 and~12. As in Section~\ref{sec:powerchord}, some harmonics above the 8\textsuperscript{th} appear to be resolved.

\subsection{Heinz Holliger, multiphonics study for oboe}\label{sec:holliger}

This section examines the multiphonic that opens Holliger's \textit{Studie \"uber Mehrkl\"ange} for solo oboe \citep{holliger1971studie}. The resulting tone lies near the boundary between quasi-harmonic and inharmonic: while the partials exhibit a quasi-harmonic progression, the perceived pitches do not align with a harmonic series. The score specifies a ``chord'' (\textit{Mehrkl\"ange}) to be produced using B\ensuremath{\flat}3 fingering with the half-hole key partially opened. No target pitches are indicated.

%\vspace{-.15cm}

\begin{figure}[h!]
\includegraphics[width=\textwidth]{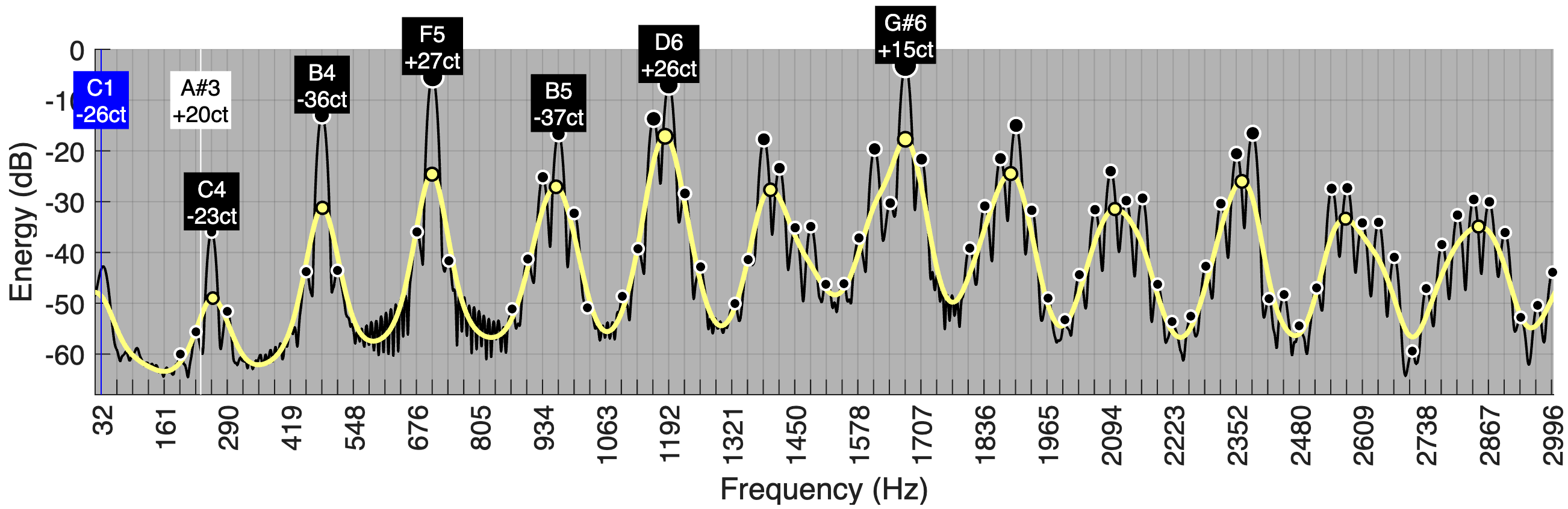}
\caption{Heinz Holliger, \textit{Studie \"uber Mehrkl\"ange f\"ur Oboe Solo}, first ``note'', weighted audio. Black, power spectrum. Yellow, power spectrum, smoothed. Blue box, predicted pitch value according to temporal modeling, of which the x-axis ticks are multiples. White box, same value but for the smoothed spectrum.} \label{fig:holliger}
\end{figure}

\begin{figure}[h!]
\includegraphics[width=\textwidth]{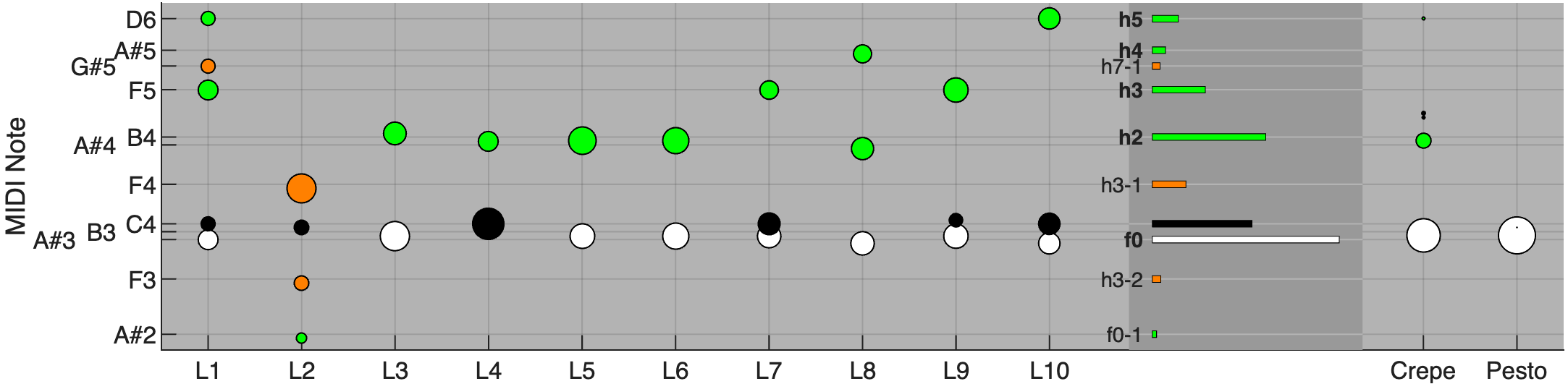}
\caption{Holliger, \textit{Studie \"uber Mehrkl\"ange}, tone from Fig.~\ref{fig:holliger}, listening test results. The harmonics correspond to the $f_0$ of the carrier (smoothed power spectrum). The pitches in black correspond to individual partials that are close to, but distinct from, the $f_0$.} \label{fig:holliger_results}
\end{figure}

Fig.~\ref{fig:holliger} reveals a low-frequency quasi-periodic component (low C1, 32~Hz) that is not perceived. The spectrum's envelope is modulated by a higher-frequency component, close to A$\sharp$3 (236~Hz). This architecture recalls FM modulation, where sidebands spaced by the modulation frequency (here, C1) appear around the carrier (A$\sharp$3). The next section features an electronic FM-based tone. 

\newpage

Fig.~\ref{fig:holliger_results} shows the listening test results for this sample. The perceived pitches correspond to the carrier (A$\sharp$3 and harmonics). The modulation is only perceived through one partial, C4. CREPE and PESTO echo this perception, with CREPE showing a few octave jumps.

\subsection{Primaal, \textit{¡Fire!}, bass}\label{sec:fire}

\begin{figure}[h!]
\includegraphics[width=\textwidth]{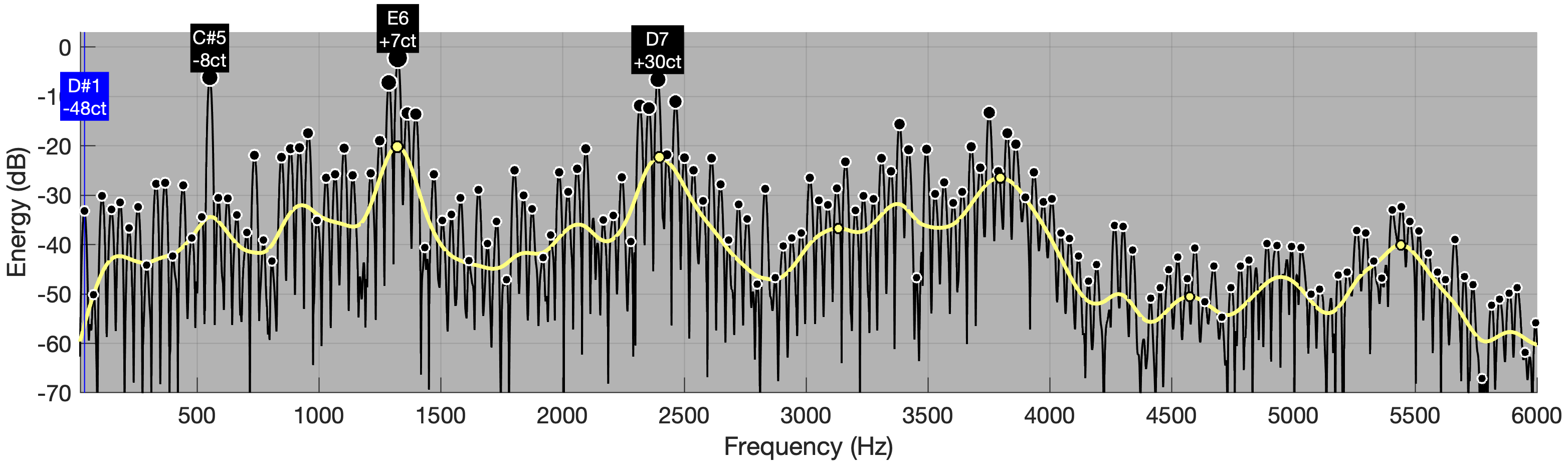}
\caption{Primaal, \textit{¡Fire!}, bass, weighted audio,
one frame. Black, power spectrum. Yellow, power spectrum, smoothed. Blue box, pitch value predicted from temporal modeling.} \label{fig:fire}
\end{figure}

This section examines a frame from a tone in the bass track of Primaal's \textit{¡Fire!}, whose production involved FM synthesis and a resonator \citep{deruty2025primaal}. As the track features continuously evolving frequencies, the sample under examination is resynthesized from a single 0.1 s frame near the start of the loop. Fig.~\ref{fig:fire} reveals a low-frequency quasi-periodic component (between D1 and D$\sharp$1, 37.8~Hz), whose spectrum is modulated by higher-frequency components that, unlike in Holliger's case, do not exhibit a harmonic relationship.

\begin{figure}[h!]
\includegraphics[width=\textwidth]{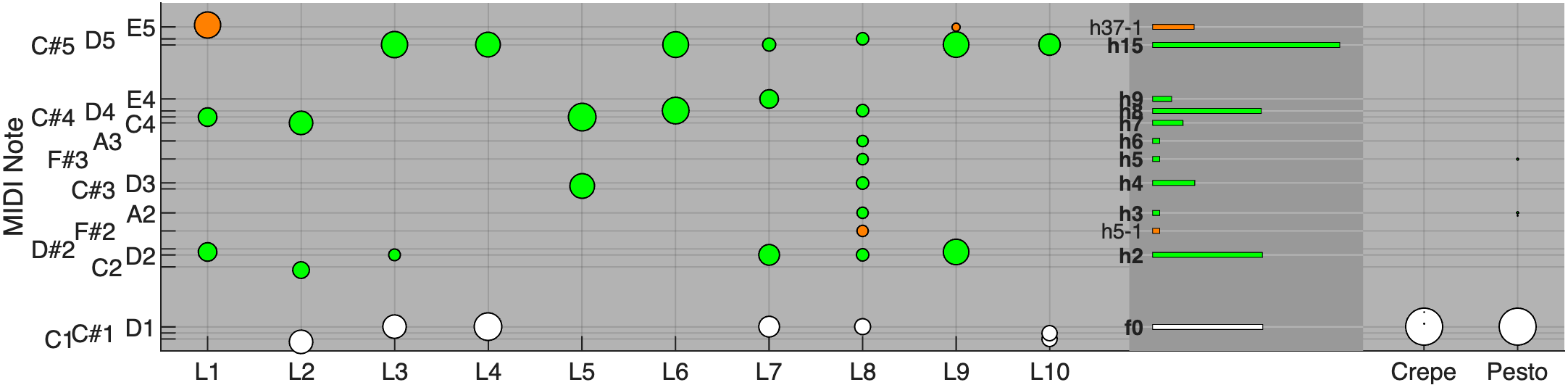}
\caption{Primaal, \textit{¡Fire!}, tone from Fig.~\ref{fig:fire}, listening test results. The association between D6 and harm. 37 (D7) stems from the high energy of this partial. Low-register pitches around C/C\#/D are merged in the bar graph.} \label{fig:fire_results}
\end{figure}

Fig.~\ref{fig:fire_results} shows the listening test results for this sample. C1 is conflated with D1 and taken as corresponding to the pitch value predicted by temporal modeling (D\#1). Such broad quantization is justified by the limited frequency resolution of the auditory system at low frequencies \citep{moore1990auditory}. For clarity of representation, the same procedure is applied to C2/D2 and C\#3/D3, even though the resulting associations with the signal are questionable. The perception of high upper partials (15 and 37) as individual pitches likely derives from the high loudness of the corresponding partials.

\section{Inharmonic tones}\label{sec:inharm}

In this section, we consider tones that are inharmonic according to the criteria in Section~\ref{sec:background}. Pitches to be transmitted are specified in one case only (Section~\ref{sec:piano}).

\subsection{Primaal, \textit{Silver}, bass}\label{sec:silver}

\begin{figure}[h!]
\includegraphics[width=\textwidth]{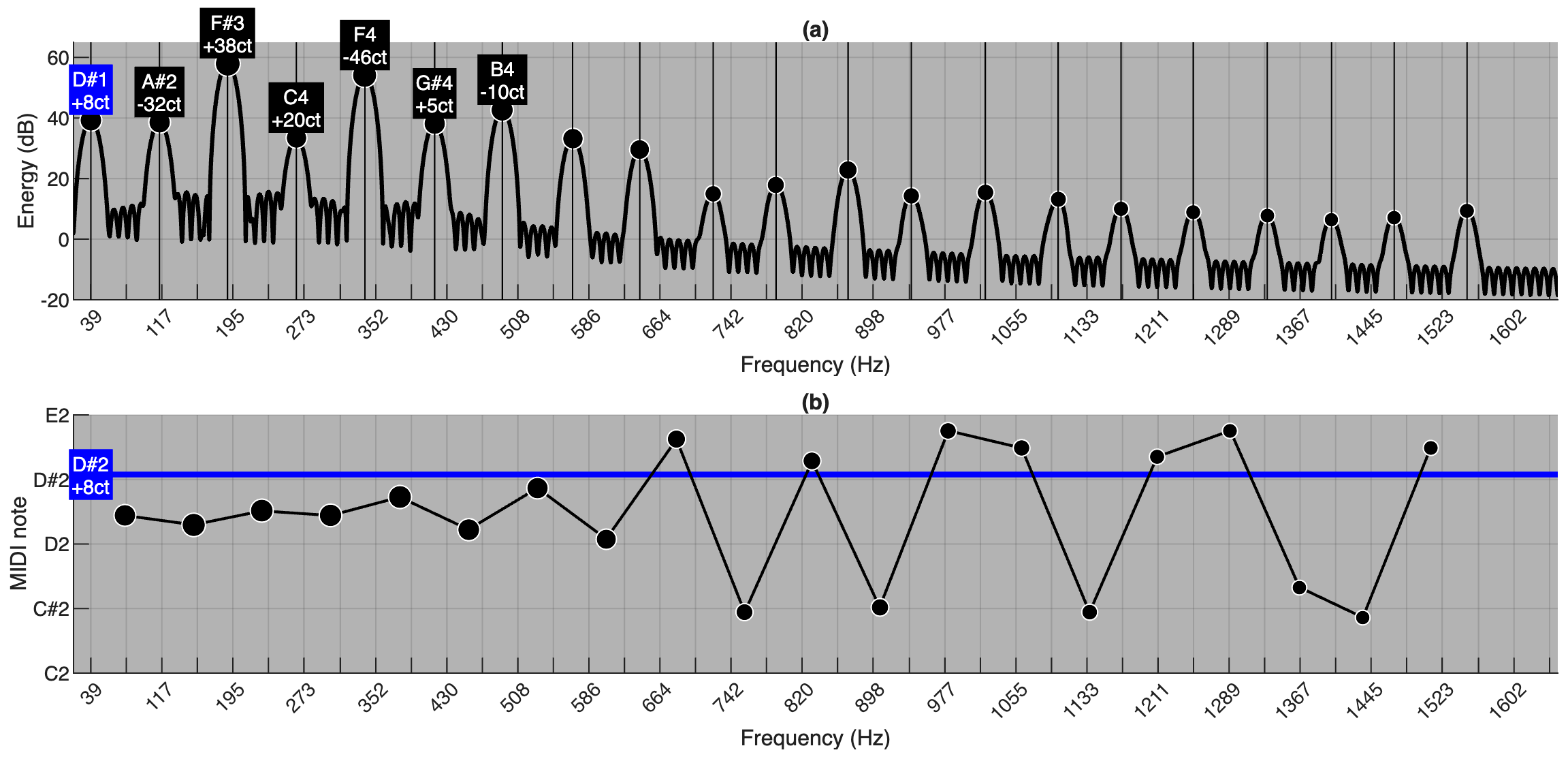}
\caption{Primaal, \textit{Silver}, bass, resynthesized bass frame, weighted audio. (a) Power spectrum. The x-axis ticks are multiples of the first peak's frequency (in blue). (b) Frequency differences between consecutive partials. The blue line denotes twice the frequency of the first peak.}
\label{fig:silver}
\end{figure}

This example derives from the bass track in Primaal's \textit{Silver} \citep{primaal2023silver}, produced using Omnisphere's `808 Woofer Warfare' patch \citep{deruty2025primaal}. As the bass track features continuously evolving frequencies, the sample under examination is resynthesized from a single 0.1~s frame in a relatively stable part. Fig.~\ref{fig:silver} shows the inharmonic character of the tone. In particular, the frequency differences between consecutive partials are centered at a lower frequency than that of the first partial. Because only odd harmonics are present, the first partial and the differences between consecutive partials differ by one octave.

\begin{figure}[h!]
\includegraphics[width=\textwidth]{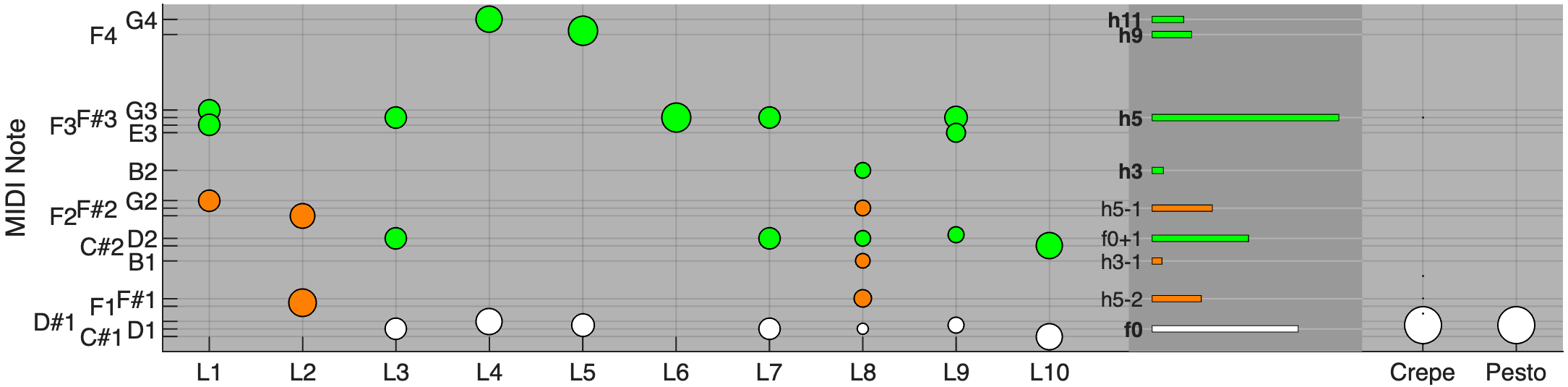}
\caption{Primaal, \textit{Silver}, bass, tone from Fig.~\ref{fig:silver}, listening test results. Pitches around F\# are merged in the bar graph. D2 is expressed as ``$f0$+1'' as the tone features only odd harmonics.}
\label{fig:silver_results}
\end{figure}

Fig.~\ref{fig:silver_results} shows the listening test results for this sample. Listeners report the perception of the $f_0$, some harmonics, along with some downward transpositions. CREPE and PESTO follow the $f_0$.

\subsection{Walter's piano multiphonics}\label{sec:piano}

\begin{figure}[h!]
\includegraphics[width=\textwidth]{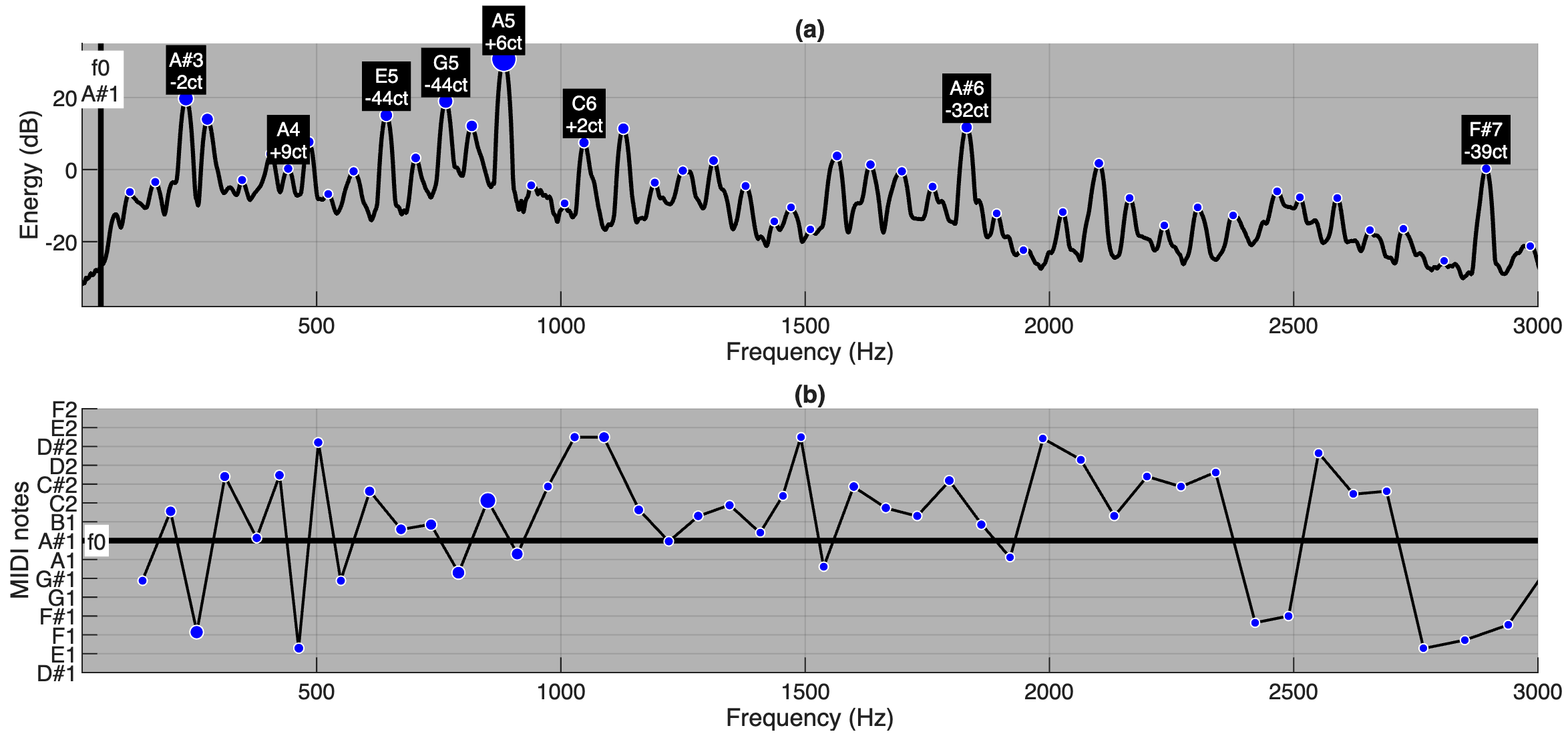}
\caption{Walter, multiphonic from \textit{Versunkene Form}, weighted audio. (a) Power spectrum. White box, predicted pitch value according to temporal modeling. (b) Frequency differences between consecutive partials.} \label{fig:walter}
\end{figure}

The work of Caspar Johannes Walter \citeyearpar{walter2020multiphonics} involves pure multiphonics on vibrating strings. While his technique is similar to Fallowfield's, the resulting tones can be strongly inharmonic. Fig.~\ref{fig:walter} shows the signal analysis of a multiphonic from \textit{Versunkene Form} for piano (2009): the ``M2'' note before bar~15 in Fig.~8 of \citet{walter2020multiphonics}. The tone exhibits sufficient periodicity to allow an $f_0$ to be inferred from temporal modeling. This $f_0$ is low (A\#1, ca.~60\,Hz), and the corresponding partial is missing. As shown in Fig.~\ref{fig:walter}(b), the partial positions display marked inharmonicity.

\begin{figure}[h!]
\includegraphics[width=\textwidth]{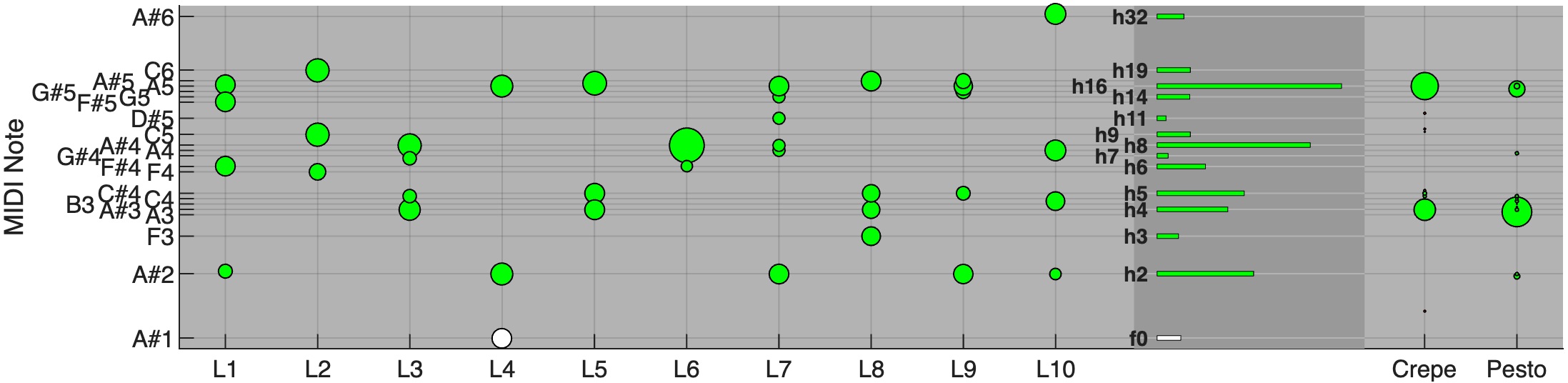}
\caption{Walter, \textit{Versunkene Form}, tone from Fig.~\ref{fig:walter}, listening test results. } \label{fig:walter_results}
\end{figure}

Fig.~\ref{fig:walter_results} shows the listening test results for this sample. Despite the high degree of inharmonicity, listeners report pitches that follow a harmonic progression. The perceived pitches poorly reflect the intended pitches (A\#3, G\#4, and D5). Both CREPE and PESTO converge on the first loud partial.

\subsection{Vitalic, \textit{No Fun}, main synthesizer part}\label{sec:nofun}

\begin{figure}[h!]
\includegraphics[width=\textwidth]{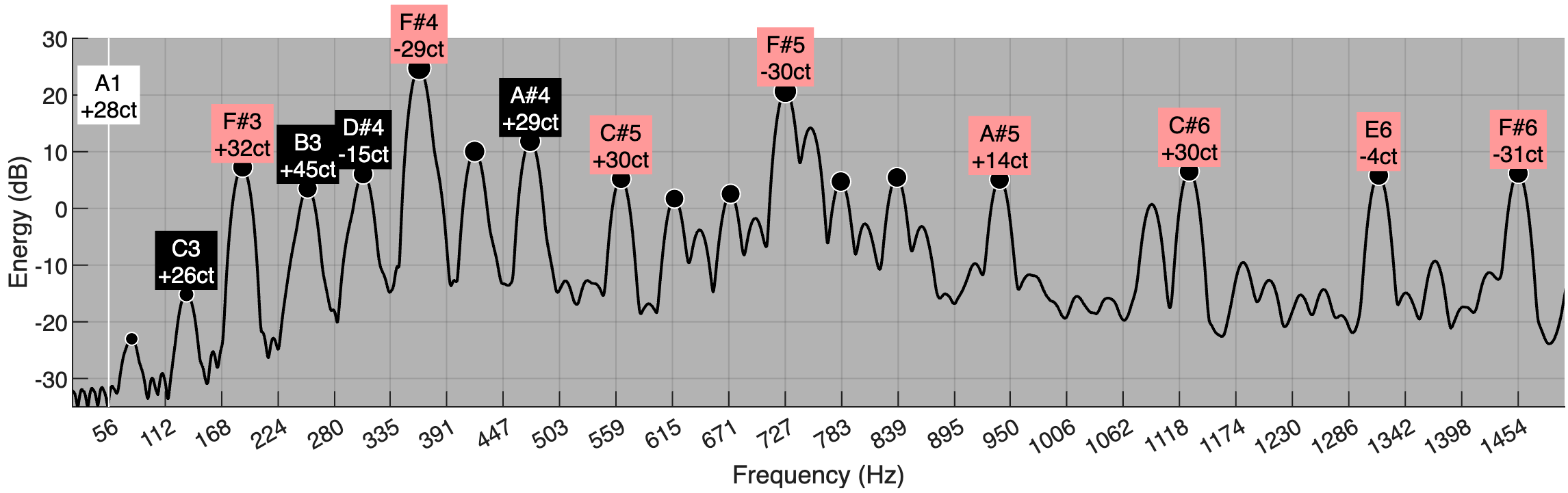}
\caption{Vitalic, \textit{No Fun}, main synthesizer part, last ``note'' of the main pattern, resynthesized sample, weighted audio. Power spectrum. White box, the predicted pitch value according to temporal modeling, of which the x-axis ticks are multiples. The partials underlined in pink form a quasi-harmonic tone.} \label{fig:nofun}
\end{figure}

A sequence from the main synthesizer part in Vitalic's \textit{No Fun} (2005) \citep{vitalic2005nofun} has been studied in \citet{deruty2025multiple}, and elaborated on in \citet{deruty2025nofun}. The present section focuses on the last ``note'' of the sequence. As it is short, the sample under examination is resynthesized from the original spectrum. Fig.~\ref{fig:nofun} shows that the tone exhibits strong regularity, with the greatest common divisor of the differences between consecutive partials corresponding to a slightly sharp A1. However, the positions of the partials do not align with integer multiples of the corresponding frequency (56 Hz). Some salient partials form a quasi-harmonic F$\sharp$3 tone.

\vspace{.2cm}

\vspace{.2cm}

\begin{figure}[h!]
\includegraphics[width=\textwidth]{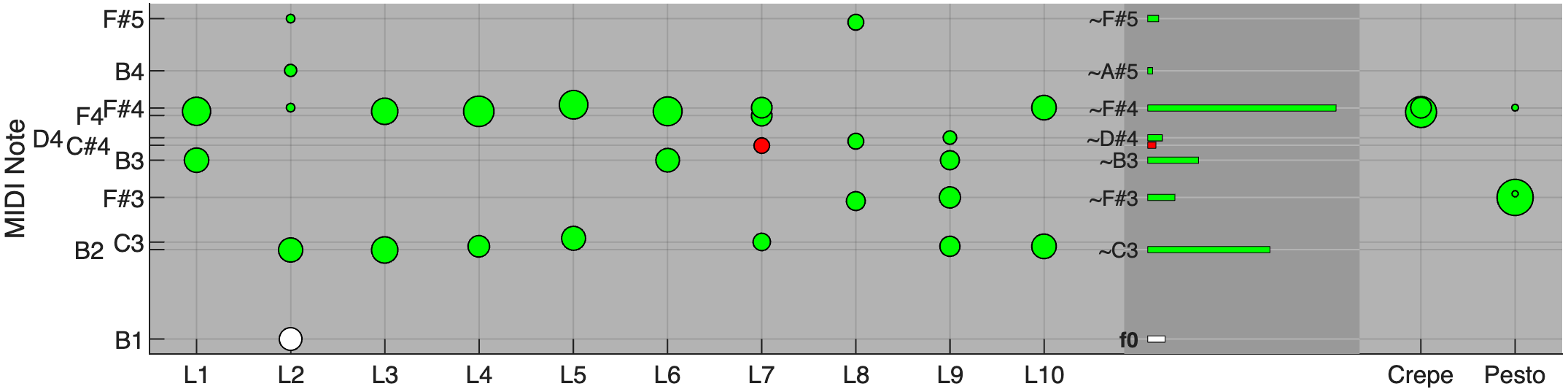}
\caption{Vitalic, \textit{No Fun}, tone from Fig.~\ref{fig:nofun}, listening test results. The approximate pitch corresponding to each partial's frequency is indicated in place of the harmonic number.
} \label{fig:nofun_results}
\end{figure}

Fig.~\ref{fig:nofun_results} shows the listening test results for this sample. Because the tone is inharmonic, pitches are referenced not to harmonic numbers but to pitches corresponding to partial frequencies. Listeners mostly appear to draw their responses from these partials. The perception of B2 in the presence of a C3\,+\,26\,ct partial suggests that this pitch arises from a subset of partials rather than from a single component -- as in Sections~\ref{sec:cellomultiphonics} and \ref{sec:Weber}. CREPE identifies the pitch as F\#4 -- the most frequently perceived pitch and the combination of the loudest partial as $f_0$ and subsequent harmonics -- whereas PESTO identifies it as F\#3, corresponding to a harmonic tone whose $f_0$ is not loud.

%% %%%%%%%%%%%%%%%%%%%%%%%%%%%%%%%%%%%%%
%% Section
%% %%%%%%%%%%%%%%%%%%%%%%%%%%%%%%%%%%%%%

%\newpage

\section{Conclusion}\label{sec:summary}

This study has shown that electronic tones used in contemporary popular music — including the widespread TR-808–style bass and single tones resulting from distorted power chords — are structurally and perceptually similar to \textit{multiphonics} in contemporary classical music. By applying the same methods to both, we demonstrate that the two classes of tones elicit multiple pitch percepts that vary across listeners. Multiphonics, it appears, are not confined to Western contemporary classical contexts but pervade contemporary popular music production.

Three specific observations emerge:
\begin{enumerate}[label=(\alph*)]
\item both tone types elicit multiple pitches that vary across listeners;
\item the perceived pitches are most often harmonically related, even for inharmonic tones, and appear to derive from subsets of salient partials; and
\item tones designed to convey a single pitch produce a weaker version of the same effect, suggesting a continuum in which conventional tones and multiphonics differ in degree of perceptual consensus rather than in kind.
\end{enumerate}

The behavior of monophonic pitch trackers in the presence of the tones considered here supports this view: octave jumps in their output may reflect intrinsic signal properties -- namely, the capacity to convey multiple pitches simultaneously -- rather than algorithmic shortcomings.

A remaining question is the prevalence of multiphonic tones in popular music, though the ubiquity of power chords alone suggests it may be high. Complementing earlier publications \citep{deruty2025emerging,deruty2025vitalictemperament,deruty2025multiple,deruty2025primaal}, this study underscores that pitch ambiguity -- at the intersection of composition, production, and perception -- may not be a marginal phenomenon, but a widespread feature of contemporary music making.

%%%%%%%%%%%%%%%%%%%%%%%%%%%%%%%%%%%%%%%%%%%%%%%%%%%%%%%%%%%%%%%%%%%%%%%%%%%%%%%%
% Bibliography
%%%%%%%%%%%%%%%%%%%%%%%%%%%%%%%%%%%%%%%%%%%%%%%%%%%%%%%%%%%%%%%%%%%%%%%%%%%%%%%%

\newpage

% For bibtex users:
\bibliographystyle{apalike}
\bibliography{mybib.bib}
\end{document}